\documentclass[conference]{IEEEtran}
\IEEEoverridecommandlockouts
\usepackage{cite}
\usepackage{amsmath,amssymb,amsfonts}
\usepackage{algorithmic}
\usepackage[titlenumbered,ruled]{algorithm2e}
\usepackage{graphicx}
\usepackage{textcomp}
\usepackage{amsmath,array,graphicx}
\usepackage{kantlipsum}
\usepackage{lettrine}
\usepackage{amsmath}
\usepackage{mathtools}
\usepackage{multirow}
\usepackage{mathtools, cuted}
\usepackage{lipsum, color}
\usepackage{amsmath}
\usepackage{subcaption}
\usepackage{graphicx}
\usepackage[font=footnotesize]{caption}

\def\BibTeX{{\rm B\kern-.05em{\sc i\kern-.025em b}\kern-.08em
    T\kern-.1667em\lower.7ex\hbox{E}\kern-.125emX}}
\begin{document}
\bstctlcite{IEEEexample:BSTcontrol}
\title{ Accelerating Beam Sweeping in mmWave Standalone 5G New Radios using Recurrent Neural Networks 
}

\author{\IEEEauthorblockN{Asim Mazin, Mohamed Elkourdi, and Richard D. Gitlin,\textit{ Life Fellow, IEEE} }
\IEEEauthorblockA{{Innovation in Wireless Information Networking Lab (\textit{i}WINLAB)
} \\
{Department of Electrical Engineering, University of South Florida},
Tampa, Florida 33620, USA \\
Email:\{ asimmazin,elkourdi\}@mail.usf.edu, richgitlin@usf.edu }
}

\maketitle

\begin{abstract}
Millimeter wave (mmWave) is a key technology to support high data rate demands for 5G applications. Highly directional transmissions are crucial at these frequencies to compensate for high isotropic pathloss. This reliance on directional beamforming, however, makes the cell discovery (cell search) challenging since both base station (gNB) and user equipment (UE) jointly perform a search over angular space to locate potential beams to initiate communication. In the cell discovery phase, sequential beam sweeping is performed through the angular coverage region in order to transmit synchronization signals. The sweeping pattern can either be a linear rotation or a hopping pattern that makes use of additional information. This paper proposes beam sweeping pattern prediction, based on the dynamic distribution of user traffic, using a form of recurrent neural networks (RNNs) called Gated Recurrent Unit (GRU). The spatial distribution of users is inferred from data in call detail records (CDRs) of the cellular network. Results show that the user’s spatial distribution and their approximate location (direction) can be accurately predicted based on CDRs data using GRU, which is then used to calculate the sweeping pattern in the angular domain during cell search.

\end{abstract}
\par 
\section{Introduction}  
\lettrine[findent=2pt]{{\textbf{M}}}{ }ilimeter wave (mmWave) is an enabling technology for 5G high data rate use cases due to the available bandwidth at these frequencies. However, the initial access in mmWave cellular systems is challenging compare to the current LTE system for two reasons. First, due to the high isotropic path-loss the mmWave communications requires high directional transmission. But the UE and gNB do not know in which directions to transmit (receive) during the initial access. Second, since the mmWave link is vulnerable to blocking and beam misalignment, more frequent initial access needs to be performed\cite{b5},\cite{b6},\cite{b7},\cite{b8}. The IEEE 802.11ad standard adopted two levels initial beamforming training for 60 GHz, where a coarser sector level sweep phase is followed by an optional beam refinement phase\cite{b12}.
\par 
Recently context information (e.g., vehicle's position) and past beam measurements stored in a database (maintained in the road side unit in vehicular communications) has been used as a hint to determine potential beam pairs\cite{b13}. Generally speaking, the initial access procedure can be improved by richer information, e.g., terminal positions, channel gain predictions, user spatial distribution, antenna configurations successfully used in previous accesses, and so on. 
\par 
The main contribution of this paper is to leverage intelligence from call detail records (CDR) data to rapidly determine the sweeping direction pattern during the cell discovery phase in mmWave cellular system using Recurrent Neural Networks (RNN) to predict the evolution of the CDR pattern.     
\par
The remainder of this paper is organized as follows. Section II discusses the initial access in standalone 5G NR, Section III presents initial access based on a machine learning approach using Recurrent Neural Networks (RNNs), Section IV results are discussed and in Section V the paper discussed the conclusions.

\section{Initial Access in standalone 5G New Radio}
The initial access in 5G New Radio (NR) standalone millimeter wave is a time-consuming search to determine suitable directions of transmission and reception. The overall idea of the envisioned mmWave initial access procedure is summarized in Fig.1 (a) \cite{b11}. The problem of interest in this paper is the cell discovery. In the cell discovery phase, one approach is sequential beam sweeping by the base station that requires a brute force search through many beam-pair combinations between the user equipment (UE) and the gNB (5G base station) to find the optimum beam-pair (i.e. the one with the highest reference received signal power (RSRP) level as shown in Fig.1 (b).The sequential search may result in a large access delay and low initial access efficiency. This paper proposes beam sweeping pattern prediction to determine the beam hopping sequence, based on the dynamic distribution of user traffic (i.e., CDRs). This is done by using a form of recurrent neural networks (RNNs) called Gated Recurrent Unit (GRU). It is worth mentioning that a UE does not only need to carry out cell search at power-up, but to support mobility. It also needs to continuously search for, synchronize to neighboring cells and estimate their reception quality. The reception quality of the neighboring cells, in relation to the reception quality of the current cell, is then evaluated to determine if a handover (for devices in RRC CONNECTED) or cell reselection (for devices in RRC IDLE) should be carried out \cite{b10}. 
\par
The standalone mmWave system is subject to significant coverage issues if beam sweeping (directional transmission) is not applied during cell search. In the current LTE system, the initial access is performed on omnidirectional channels, whereas the beamforming transmission is performed after establishing the physical link \cite{b14}. On other hand, to cope with the converge issue resulting from the increased isotropic path loss in mmWave frequencies, in 5G standalone mmWave cellular systems, the initial access must be performed on directional channels\cite{b9}. In the ongoing 5G NR standalone mmWave standards meeting, the so-called synchronization signal block (SSB) was introduced, which comprises a primary synchronization signal (PSS), a secondary synchronization signal (SSS), and a physical broadcast channel (PBCH). The synchronization signal burst was allocated 250 micro seconds, which was further divided into 14 SSB as illustrated in Fig. 2. The gNB may sweep 14 different directions (per antenna port) for the sync transmission. The exact choice of the sweeping pattern can be left to the cells; this pattern should occur periodically and the maximum periodicity must be known by the UE \cite{b9}. 
\begin{figure}[t]
\centering
\begin{subfigure}[t]{0.5\textwidth}
   \includegraphics[width=1\linewidth]{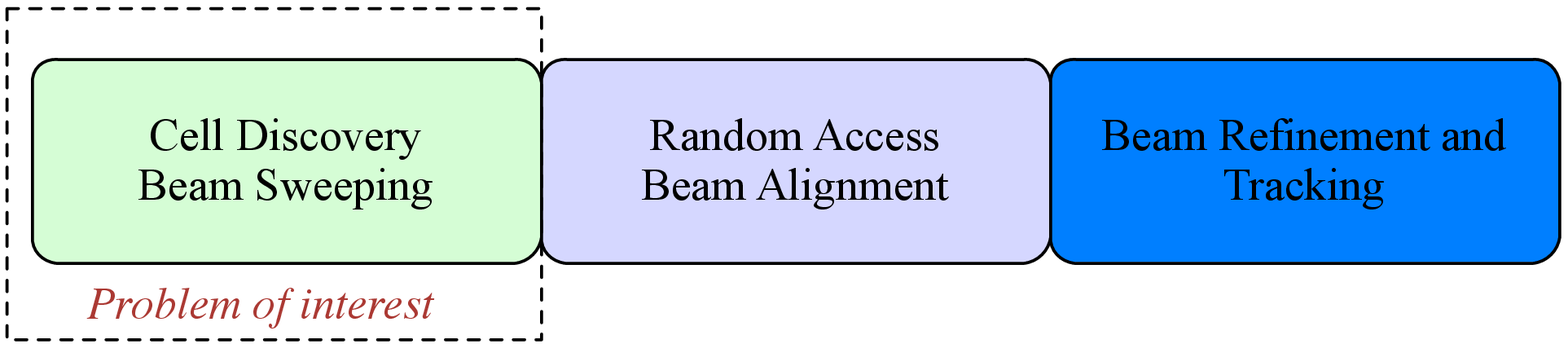}
   \caption{}
\end{subfigure}
\begin{subfigure}[t]{0.5\textwidth}
   \includegraphics[width=1\linewidth]{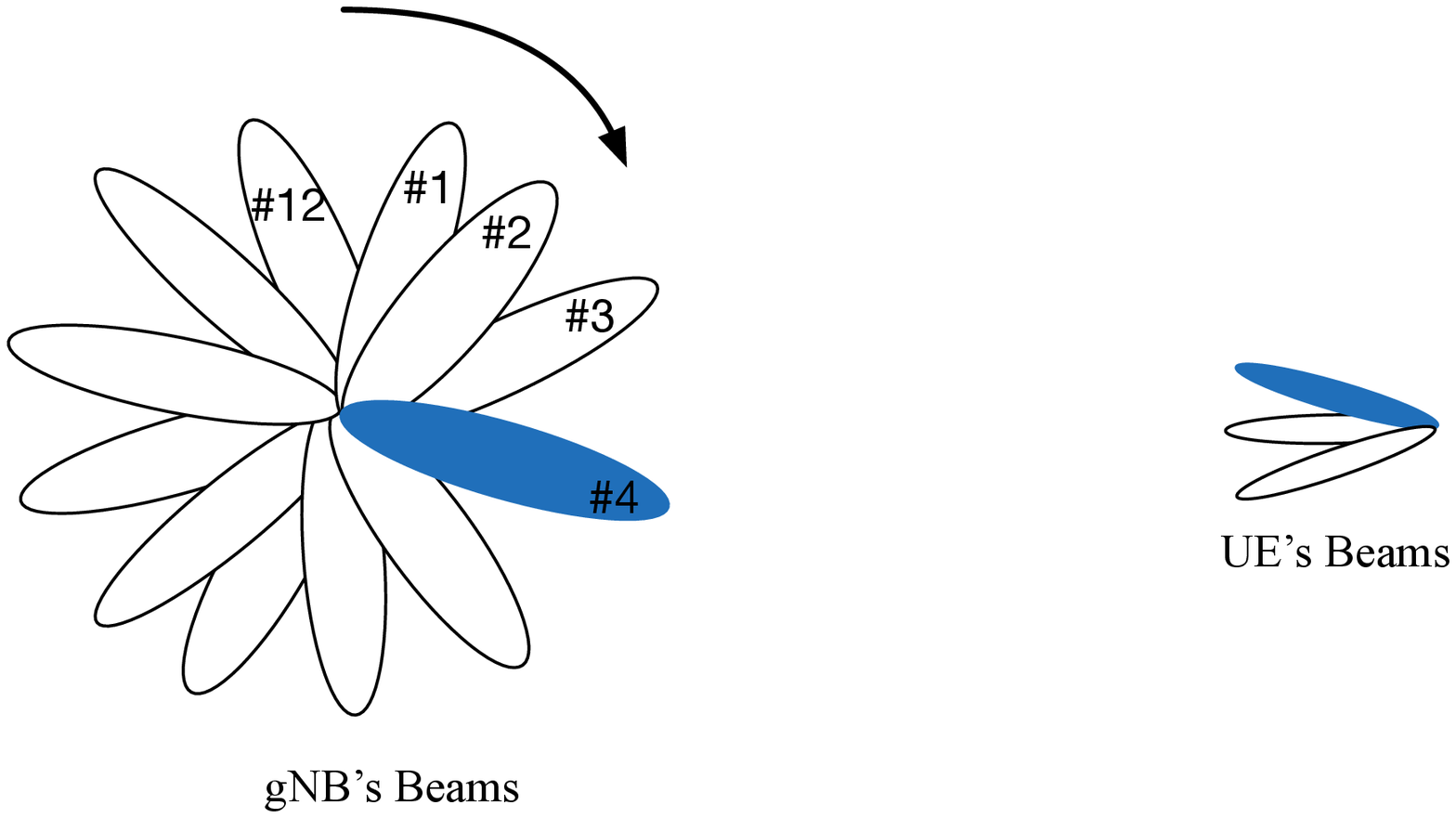}
   \caption{}
\end{subfigure}
\caption{(a) The envisioned procedure for mm-wave initial access. (b) Beam sweeping during initial access}
\end{figure}
\begin{figure}[t]
\centering
\includegraphics[width=.5\textwidth]{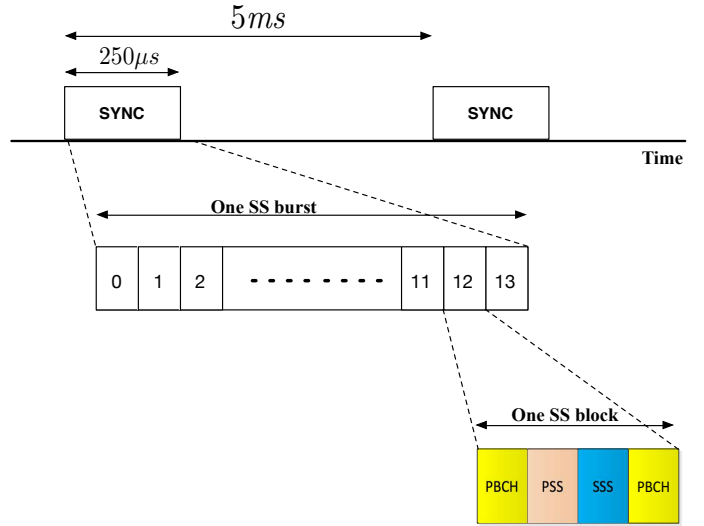}
\caption{Resources allocated to sync transmission}
\end{figure}

\section{Initial Access Machine learning approach}
To ensure that users can be quickly accessed, a form of machine learning can be used to optimize the sweeping pattern of the gNB, including beam direction and sweeping order according to the predicted user's spatial distribution from user’s historical data (e.g. delay access, access success rate and beam direction etc.)\cite{b15}. The focus of this paper is the sweeping order in the cell discovery phase. The proposed approach leverages intelligence from the CDRs data collected from Milan City network, provided by Telcom Italia as part of their Big Data challenge\cite{b1}.
\subsection{Dataset}
The data used in this paper is in form of CDRs of Internet activity, calling and text messages. The dataset measures the level of interaction of the users and the cellular network by temporally aggregating CDRs in timeslots of 10 minutes. The datasets provide spatial information about the each CDR by using the Milano Grid\cite{b2} CDR data, which contains numbered squares (square ID) that are overlaid over Milan city. The data lakes the coordinates of each CDR and only provides the square ID. Therefore to achieve the objective of the proposed data driven sweeping order, we assume that a cell is made of four squares in the Milano grid and each square represents a sector (direction). In order to determine the users activity in each sector, we count the number of CDRs that was recorded in the same timestamps in a given sector. TABLE \ref{tab:table1} presents sample data points, which show the number of CDRs on "2013-11-17" at five timestamps in four sectors denoted by \textbf{A}, \textbf{B}, \textbf{C} and \textbf{D}. In this paper the pseudo-omni beam transmission is adapted i.e. the gNB transmits the synchronization signal for a longer duration with a pseudo-omni beams. The order of the beam sweeping is determined based on time series prediction using a Neural Networks as discussed in III-B.  
\begin{table}[t]
\centering
\caption {{Number of CDRs per sector } } \label{tab:table1} 
\begin{tabular}{l c c c l} 
 \hline
 \textbf{Time} & \textbf{A} & \textbf{B} & \textbf{C}  &  \textbf{D}\\ [0.5ex] 
 \hline
 2013-11-17 22:10:00 & 3 & 3 & 3 & 5 \\ 
 2013-11-17 22:20:00 & 2 & 2 & 2 & 2 \\
 2013-11-17 22:30:00  & 3 &2  &1  &2  \\
 2013-11-17 22:40:00 & 2 & 3 & 3 & 4\\
 2013-11-17 22:50:00 & 3 & 1 & 2& 5  \\
 \hline
\end{tabular}
\end{table}
\subsection{Sweeping pattern using time series prediction}
The number of CDRs in each sector is a time series as shown in TABLE \ref{tab:table1}. To determine the sweeping pattern, or order, a recurrent neural network (RNN) is used to predict the number of CDRs in all sectors, which is used to prioritize the sweeping direction accordingly. The RNN architecture is able to captures dependencies at different time scales. A Gated Recurrent Unit (GRU) Neural Network  with 512 units is used to predict the number of CDRs in all sectors. The GRU neural net was first introduced by Cho et al. \cite{b4} for a statistical machine translation task. Fig. 3 illustrates the architecture of a GRU cell. A GRU made of two gates. The first is the update gate, which controls how much of the current cell content should be updated with the new candidate state. The second is the reset gate, which rests the memory of the cell if it is closed i.e. the unit acts as if the next processed input was the first in the sequence.The state equations of the GRU are \cite{b3}
\begin{equation}
\begin{aligned}
\textmd{reset gate} &: \mathbf{r}[t]=\sigma (\mathbf{W}_r\mathbf{h}[t-1]+\mathbf{R}_r\mathbf{x}[t]+\mathbf{b}_r),\\
\textmd{current state} &:\mathbf{h^\prime}[t]=\mathbf{h}[t-1]\odot\mathbf{r}[t],\\
\textmd{candidate state } &: \mathbf{z}[t]=g (\mathbf{W}_z\mathbf{h}^\prime[t-1]+\mathbf{R}_z\mathbf{x}[t]+\mathbf{b}_z),\\
\textmd{update gate} &: \mathbf{u}[t]=\sigma (\mathbf{W}_u\mathbf{h}[t-1]+\mathbf{R}_u\mathbf{x}[t]+\mathbf{b}_u),\\
\textmd{new state} &:\mathbf{h}[t]=(1-\mathbf{u}[t])\odot \mathbf{h}[t-1]+\mathbf{u}[t]\odot\mathbf{z}[t].
\end{aligned}
\end{equation}
where, $g(\cdot)$ is non-linear function usually implemented by a hyperbolic tangent, $\sigma$ is the logistic sigmoid\footnote{The logistic sigmoid is defined as $\sigma = \frac{1}{1+e^{-x}}$}, $\mathbf{W}_r,\mathbf{W}_z,\mathbf{W}_u$ are rectangular weight matrices,that are applied to the input $\mathbf{x}[t]$ (Number of CDRs in all sectors), $\mathbf{R}_r,\mathbf{R}_z,\mathbf{R}_u$ are square matrices that define the weights of the recurrent connections$,\mathbf{b}_r,\mathbf{b}_z,\mathbf{b}_u$ are the bias vectors and $\odot$ is the Hadamard product.    
\par 
After building GRU model, a Mean Squared Error (MSE) used as the loss-function to be minimized, which measures how closely the model's output matches the true output signals.
\begin{figure}[t]
\centering
\includegraphics[width=.5\textwidth]{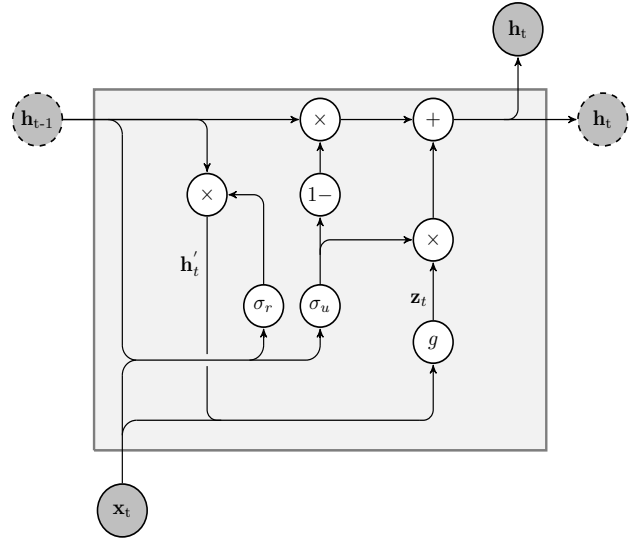}
\caption{A recurrent unit in the GRU architecture. Dark gray circles with a solid line are the variables whose content is exchanged with the input and output of the network. Dark gray circles with a dashed line represent the internal state variables, whose content is exchanged within the cells of the hidden layer. White circles with ’+’, ’−1’ and ’×’ represent linear operations\cite{b3}.}
\end{figure}

\section{Results and discussion}
In our study, two weeks of Milan CDR data (Nov 04, 2013 to
 Nov 17, 2013) were used to predict the user distribution in four sectors between the instants when the CDRs are measured. A total of 1684 sequences was used to train the GRU model and 188 to test it. The model was trained with 5 epochs, each with 50 steps. The model takes 28min 55s to train with the above mentioned epochs. Fig. 4 depicts the prediction performance of the GRU model on the tested sequences. It can be seen that the prediction are very close to the ground truth\footnote{In machine learning, the term "ground truth" refers to the accuracy of the training set's classification for supervised learning techniques.} most of the time. Based on the prediction the pseudo-omni beam can be directed toward the sector with a maximum number of CDRs. In case that the number of CDRs are equal the gNB chooses the sweeping pattern randomly. 
\begin{figure}[t]
\centering
\begin{subfigure}[ht]{0.5\textwidth}
   \includegraphics[width=1\linewidth]{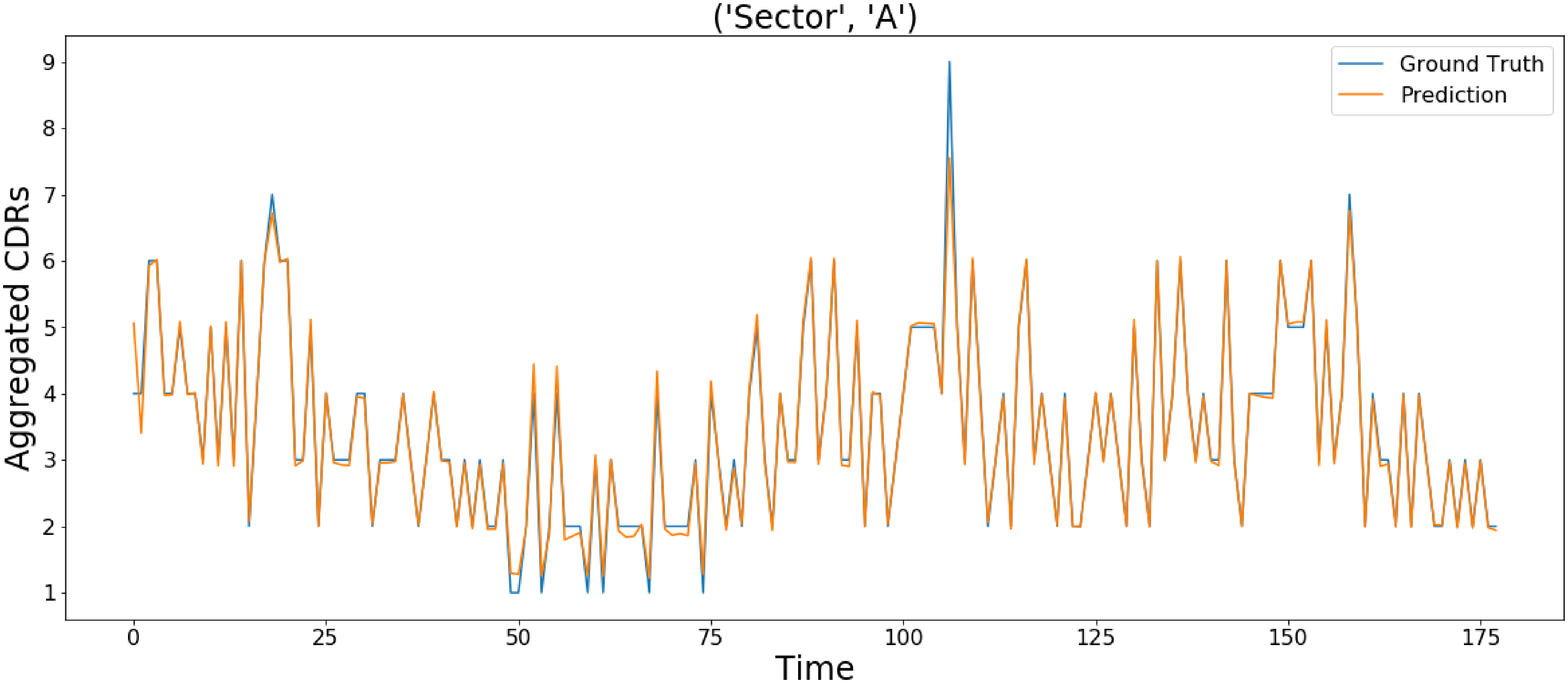}
   \caption{}
\end{subfigure}

\begin{subfigure}[ht]{0.5\textwidth}
   \includegraphics[width=1\linewidth]{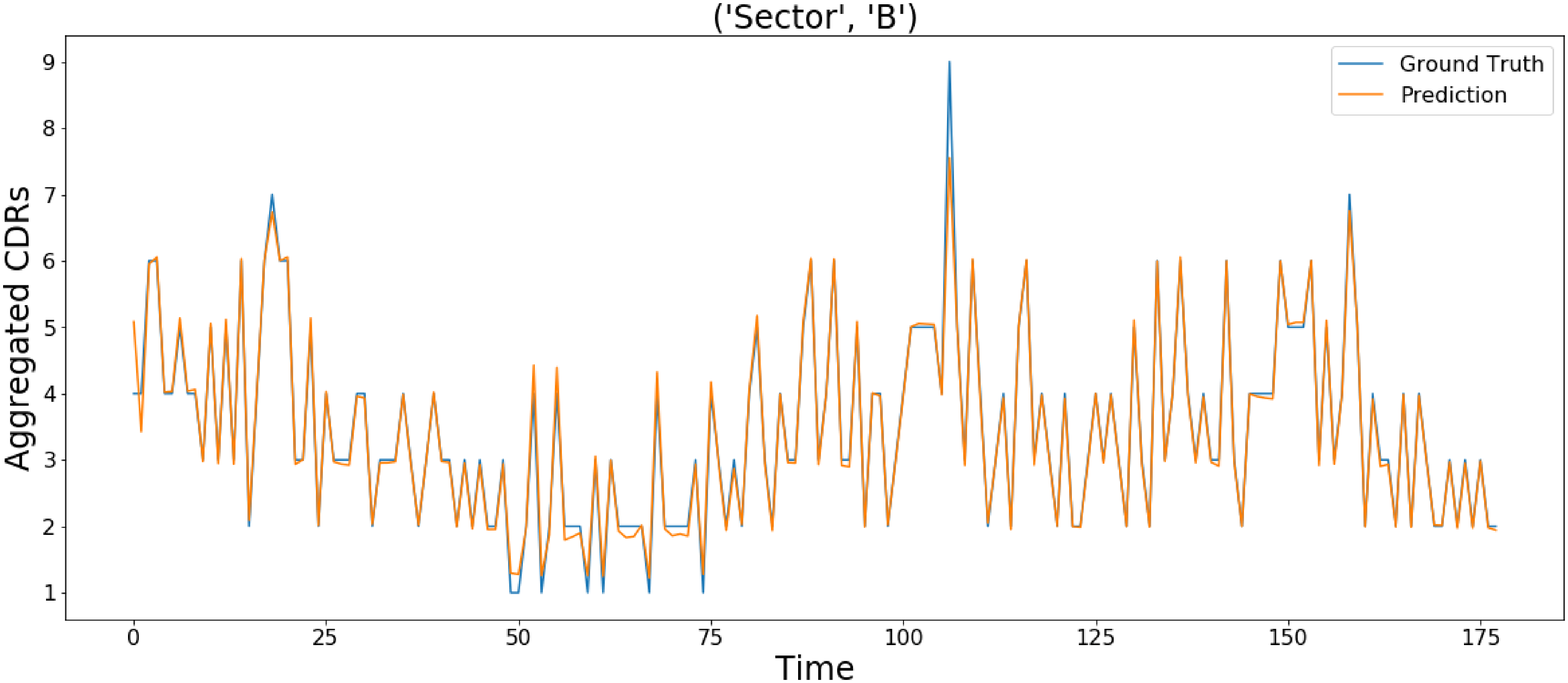}
   \caption{}
\end{subfigure}
\begin{subfigure}[ht]{0.5\textwidth}
   \includegraphics[width=1\linewidth]{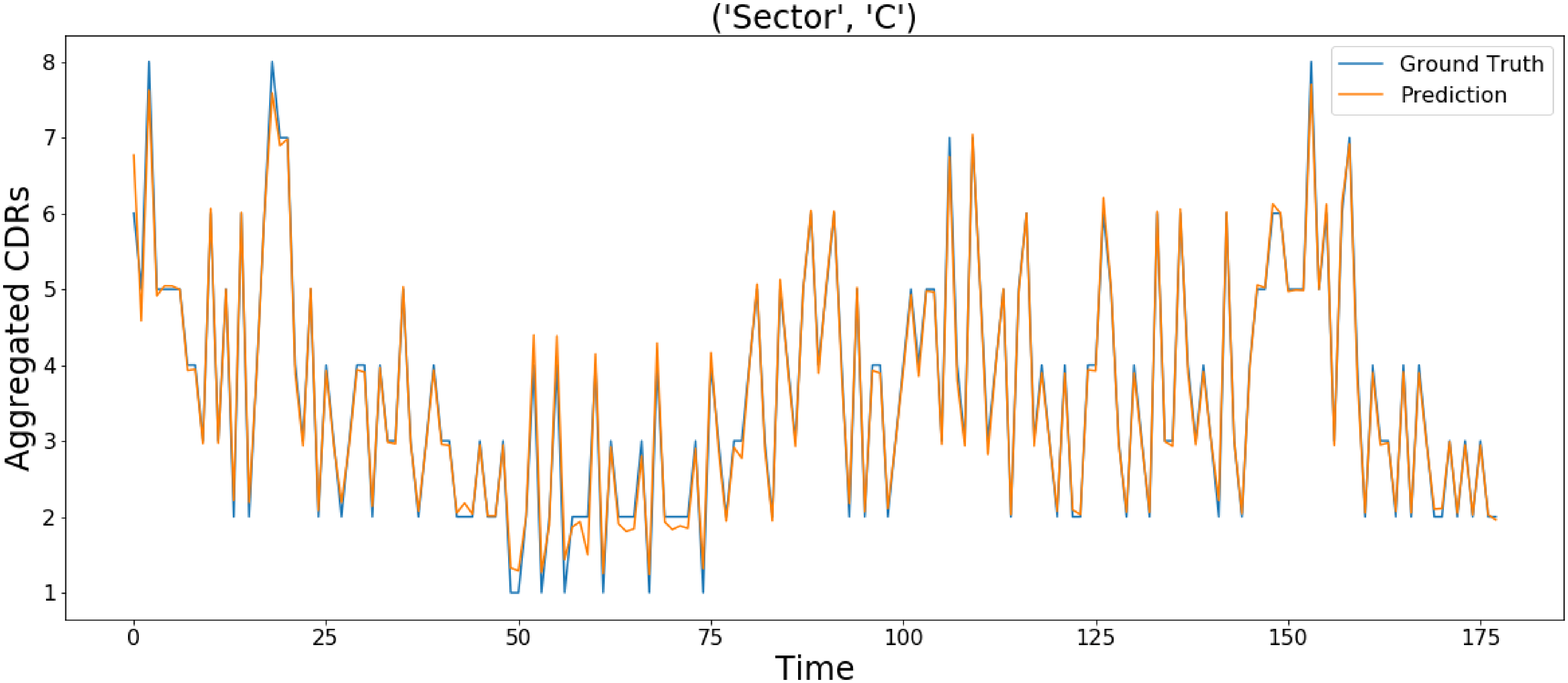}
   \caption{}
\end{subfigure}
\begin{subfigure}[t]{0.5\textwidth}
   \includegraphics[width=1\linewidth]{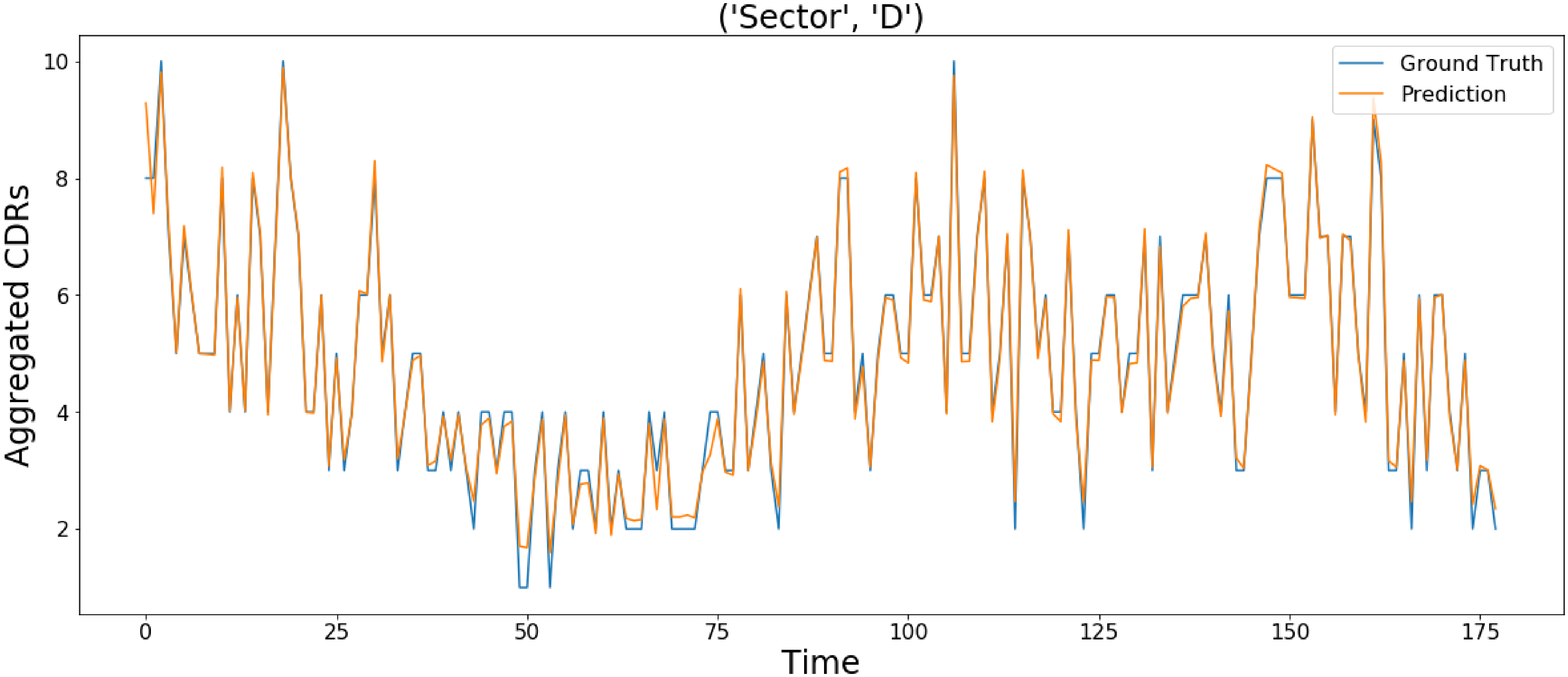}
   \caption{}
\end{subfigure}
\caption{CDRs prediction and ground truth for four sectors (a) Sector A,(b) sector B, (c) sector C and (d) sector D.}
\end{figure}

 \section{Conclusion}
Data driven beam sweeping (hopping) patterns have been introduced in this paper. It is shown that the GRU neural network can predict the CDRs with high accuracy, which is used to adjust the sweeping pattern in the angular domain. Although, the pseudo-omni was considered due to the lack of exact user location in the gNB coverage, sweeping with narrow beam can be done if the data reveals more information about the locations. Future research directions are to quantify the access delay of the data driven sweeping order using the GRU neural net and compare it with sequential sweeping.  When narrow beams are used to transmit the synchronization signals they have a higher beamforming gain than pseudo-omni beams and facilitate the selection of the best beam pair between the gNB and UE during the cell search.

\bibliographystyle{IEEEtran}
\bibliography{IEEEabrv,Reference}

\end{document}